\def\apj{\rm ApJ}
\def\apjl{\rm ApJL}
\def\apjs{\rm ApJS}
\def\aj{\rm AJ}
\def\mnras{\rm MNRAS}

\documentclass{kapproc} % Computer Modern font calls

\usepackage{procps,phase14} 

\usepackage[dvips]{graphicx}

\upperandlowercase

\kluwerbib % will produce this kind of bibliography entry:

\begin{document}

%------------ article title  ------------------->>

% If you use \\'s , please supply an alternate version of the title
% in square brackets, i.e., 
%\articletitle[Communism, Sparta, and Plato]
%{COMMUNISM, SPARTA,\\ and PLATO}

\articletitle[Galaxy Formation]{Cosmological Simulations of Galaxy \\
Formation II: Matching the Observational \\ Properties of Disk
Galaxies}

% Second method:
--------------

\author{Fabio Governato\altaffilmark{1,2}, Beth Willman\altaffilmark{3}, 
         Lucio Mayer\altaffilmark{4}}

\altaffiltext{1}{INAF-Brera, Milano. fabio@astro.washington.edu}
\altaffiltext{2}{University  of Washington, Seattle, WA}
\altaffiltext{3}{NYU, Gotham City. bw427@nyu.edu }
\altaffiltext{4}{Institute fur Astronomie, ETH Zurich, Switzerland.lucio@phys.ethz.ch }

\begin{abstract}

We used fully cosmological, high resolution N-body + SPH simulations
to follow the formation of disk galaxies with a rotational velocity
between 140 and 280Km/sec in a $\Lambda$CDM universe.  The simulations
include gas cooling, star formation (SF), the effects of a uniform UV
background and a physically motivated description of feedback from
supernovae (SN). The host dark matter (DM) halos have a spin and last major
merger redshift typical of galaxy sized halos as measured in recent
large scale N--Body simulations. Galaxies formed rotationally
supported disks with realistic exponential scale lengths and fall on
the I-band and baryonic Tully Fisher relations.  The combination of UV
background and SN feedback drastically reduced the number of visible
satellites orbiting inside a Milky Way sized halo, bringing it fair
agreement with observations.  Feedback delays SF in small galaxies and
more massive ones contain older stellar populations.  The current star
formation rates as a function of galaxy stellar mass are in good
agreement with those measured by the SDSS.

\end{abstract}

%------------ body of article ------------------->>

\section{Introduction}

N--Body/gasdynamical simulations have become the primary tools with
which to model galaxy formation in a cosmological context.  They are
necessary to follow the internal structure of galaxies as well as the
complex interplay between baryon cooling and feedback.  These studies
are of fundamental importance to answer a number of challenges faced
by the hierarchical model of structure formation: (a) Early numerical
simulations of galaxy formation reported a catastrophic loss of
angular momentum in the baryonic component, leading to the formation
of galaxies with very small disk scale lengths and dominant spheroidal
components (e.g. Navarro \& Steinmetz 2002); (b) dark matter-only simulations
predict over an order of magnitude more subhalos around Milky Way-like
galaxies than the number of dwarfs observed around the Milky Way and
M31 (Moore et al. 1998., Willman et al. 2004); (c) Observations show
that  more massive late-type galaxies have older stellar
populations than less massive late-type galaxies (Mac Arthur et
al. 2004).  This finding is at face value in contradiction with the
fact that in CDM scenarios less massive halos form first. {\it These
very different aspects of the formation and evolution of galaxies need
to be simultaneously addressed by simulations of galaxy formation}.

\begin{table*}
\centering
\begin{tabular}{cclcccc}
\hline\hline
Run & Virial Mass  & 
Vc$^{a}$ & spin & 
Last major merger  &
  $\epsilon$ &
 N$_{tot}^{b}$ at z=0 \\
${}$ & M$\odot$ & Km/sec & $\lambda$ & z & kpc & dark+gas+stars \\
\hline
DWF1 &1.6 10$^{11}$  &   70  & 0.01  & 2.2  &  0.3  & $\sim$ 860.000 \\
MW1$^1$ & 1.15 10$^{12}$  &  134    & 0.04  & 2.5  & 0.6-0.3$^1$  & $\sim$ 700.000 - 4.000.000$^2$ \\
GAL1& 3.1 10$^{12}$ &  185    & 0.035 & 2.75  & 1.  & $\sim$ 480.000  \\
\hline
\end{tabular}
\caption[Summary of the properties of the three cosmological halos]
{Summary of the properties of the three cosmological halos. $^a$:
Circular velocity at virial radius. $^b$: number of gas and star
particles changes slightly depending on $\epsilon$SN. $^{1\&2}$: smaller
softening/larger N values are for a z=0.5 ultra-high resolution run. }
\label{tbl-3}
\end{table*}

\section{Code and Simulations}

We used GASOLINE (Wadsley, Quinn \& Stadel 2004), a smooth particle
hydrodynamic (SPH), parallel treecode that is both spatially and
temporally adaptive with individual particle time steps.  The code
includes a treatment of i) Compton and radiative cooling, ii) star
formation and a supernova (SN) feedback and iii) a UV background from
QSOs and star forming galaxies following an updated version of the
Haardt \& Madau model predictions (1996, Haardt 2005, private
communication). The facts that supernova feedback and the UV
background can reduce gas retention and accretion in halos with low
virial temperature makes them possible solutions to the problems
mentioned in the introduction (e.g. Benson et al. 2002).  The current
implementation of star formation and feedback from stellar process is
described in detail in Stinson et al (2005). Feedback follows
qualitatively the algorithm implemented by Thacker \& Couchman (2000).
We assume that the energy released into the ISM stops the gas from
cooling and forming stars over some timescale, thereby the gas will
just expand adiabatically.  The time scale for the cooling shutoff and
the amount of mass affected are now physically motivated and chosen
following McKee \& Ostriker (1977). The only two free parameters in
our algorithm are star formation efficiency and the fraction of SN
energy dumped into the ISM.  These parameters are tuned to reproduce
the properties present day disk galaxies (Governato et al. 2005) and
are then applied to cosmological simulations. The SN efficiency
parameter ($\epsilon$SN) was varied from 0.2 to 0.6. Best overall
results were obtained with $\epsilon$SN = 0.4 to which all plots refer
unless otherwise noted.  A Miller Scalo IMF is assumed. Cosmological
galaxies were simulated in a $\Lambda$CDM concordance cosmology using
the ``renormalization'' technique to maximize the resolution in the
region of interest (Katz \& White 1993).  {\it Three DM halos were
selected with the criteria of a) a last major merger (defined as a 3:1
mass ratio) at redshift 2.2 $<$ z $<$ 2.75, and b) no halos of similar
or larger mass within a few virial radii}.  The physical and numerical
parameters of our runs are summarized in Table 1.

The last major merger redshift (z$_{lmm}$) range is typical for galaxy
sized DM halos in a concordance cosmology, as recently shown in large
N-body simulations by Li, Mo \& Van den Bosch (2005) and Maller et al
(2005).  Global magnitudes for our models were obtained coupling the
star formation histories (SFH) of our simulations with GRASIL (Silva
et al. 1998), a code to compute the spectral evolution of stellar
systems taking into account the effects of dust reprocessing.

\begin{figure}[t]
\plottwo{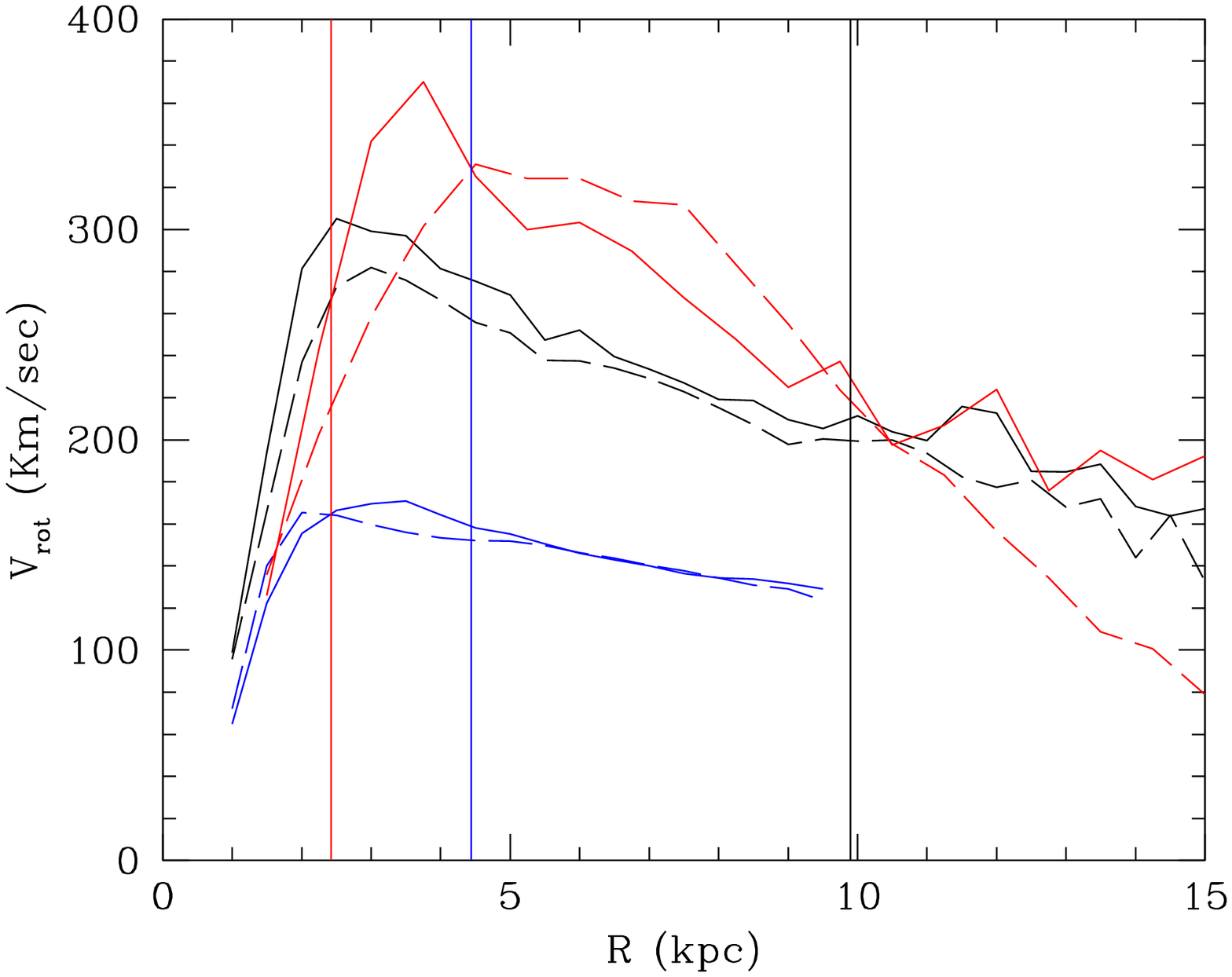}{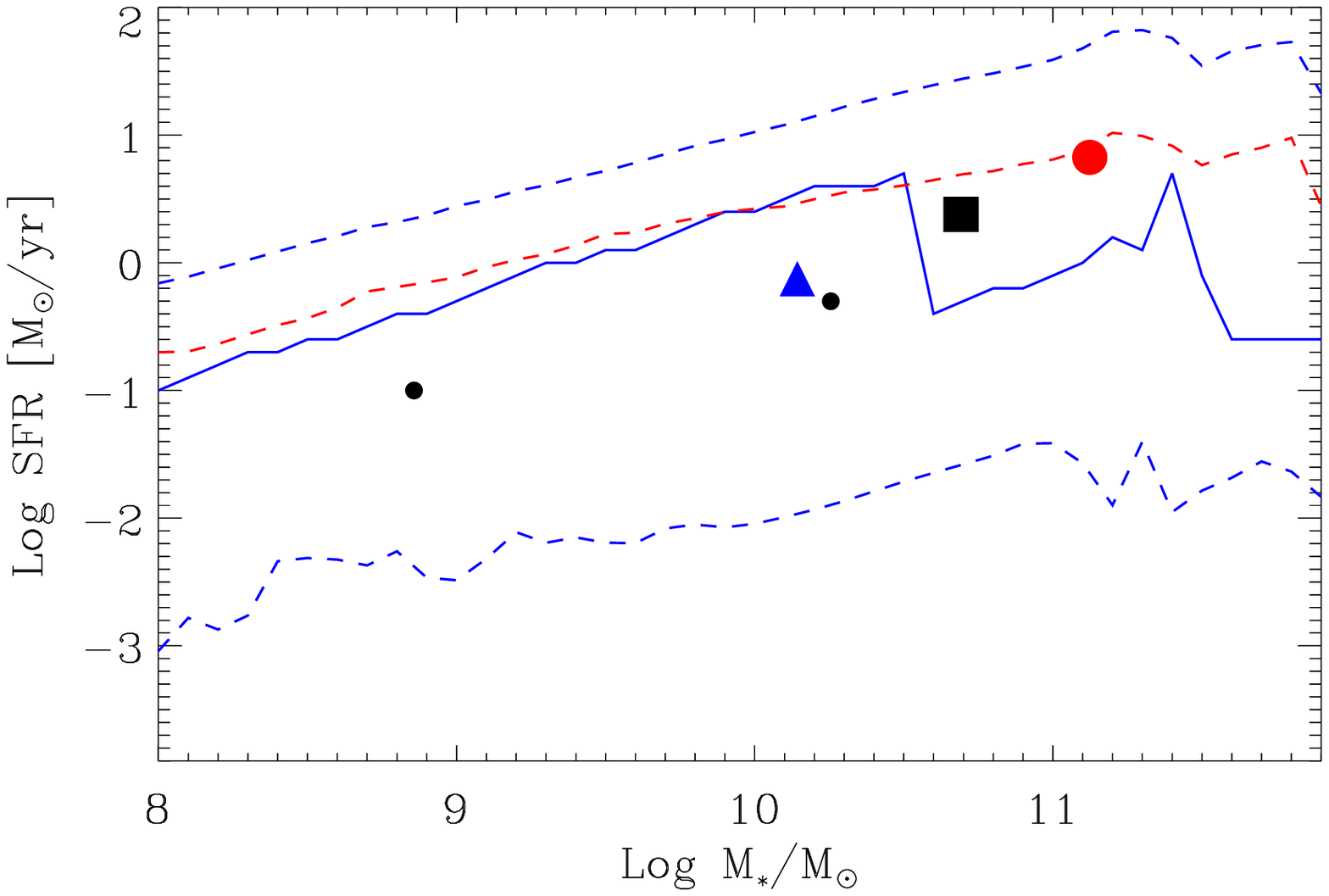}
\caption{Left Panel: Azimuthally averaged
 rotation velocity of cold gas (solid line) and stars younger than 6.5
 Gyrs (dashed line). From top to bottom: GAL1,MW1,DWF1. The vertical
 lines show 2.2 disk scale lengths for each simulated galaxy, the
 point at which the rotation velocity V$_{TF}$ is measured. Right
 panel: The present day star formation rate of our simulated galaxies
  vs their stellar mass compared with the SDSS sample in
 Brinchmann et al. 2005. (blue triangle: DWF1, solid square: MW1, red
 circle: GAL1, solid dots: simulated small galaxies within the high
 res region of MW1.}
\label{twobarrl}
\end{figure}

\begin{figure}[t]
\plottwo{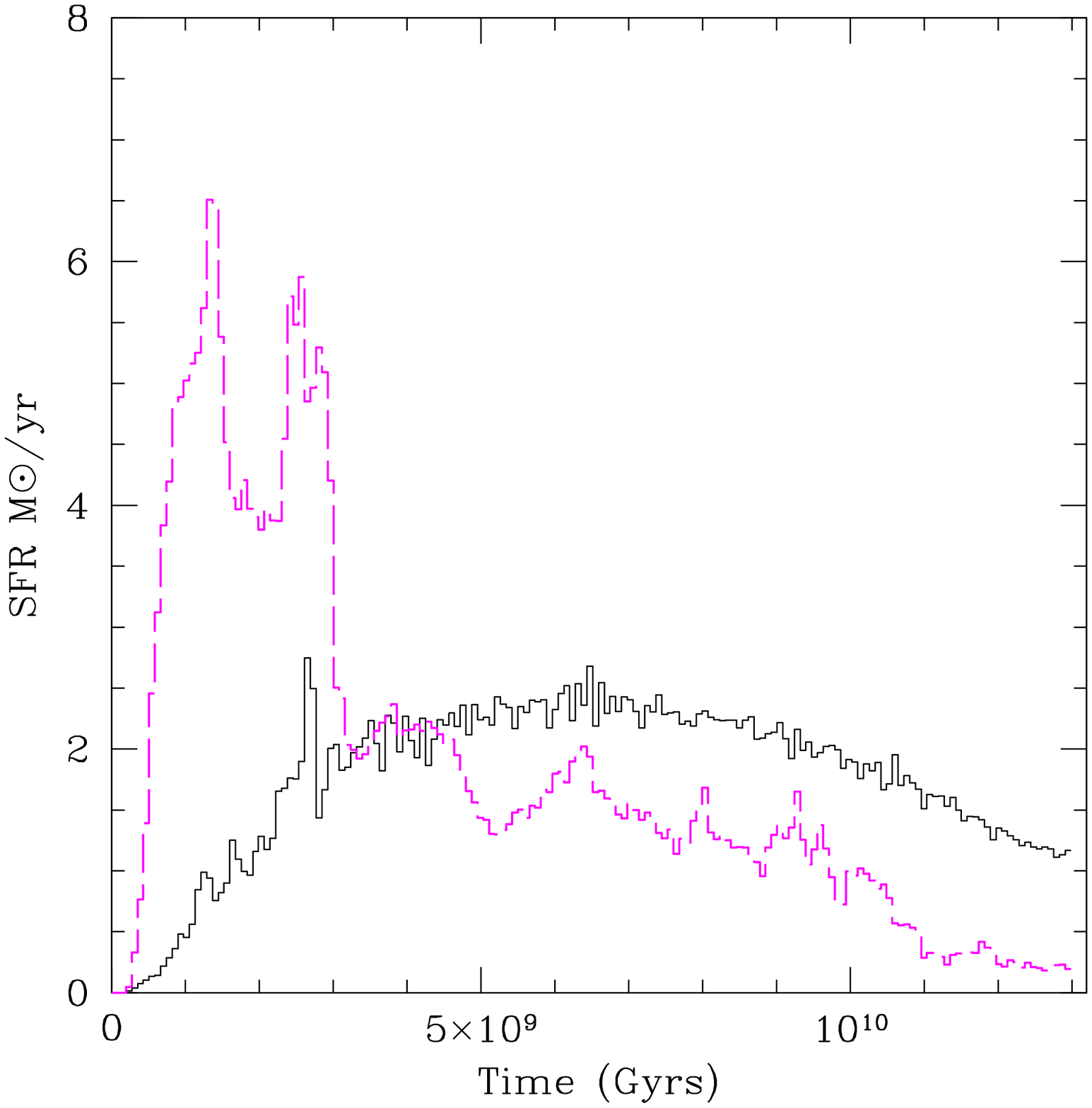}{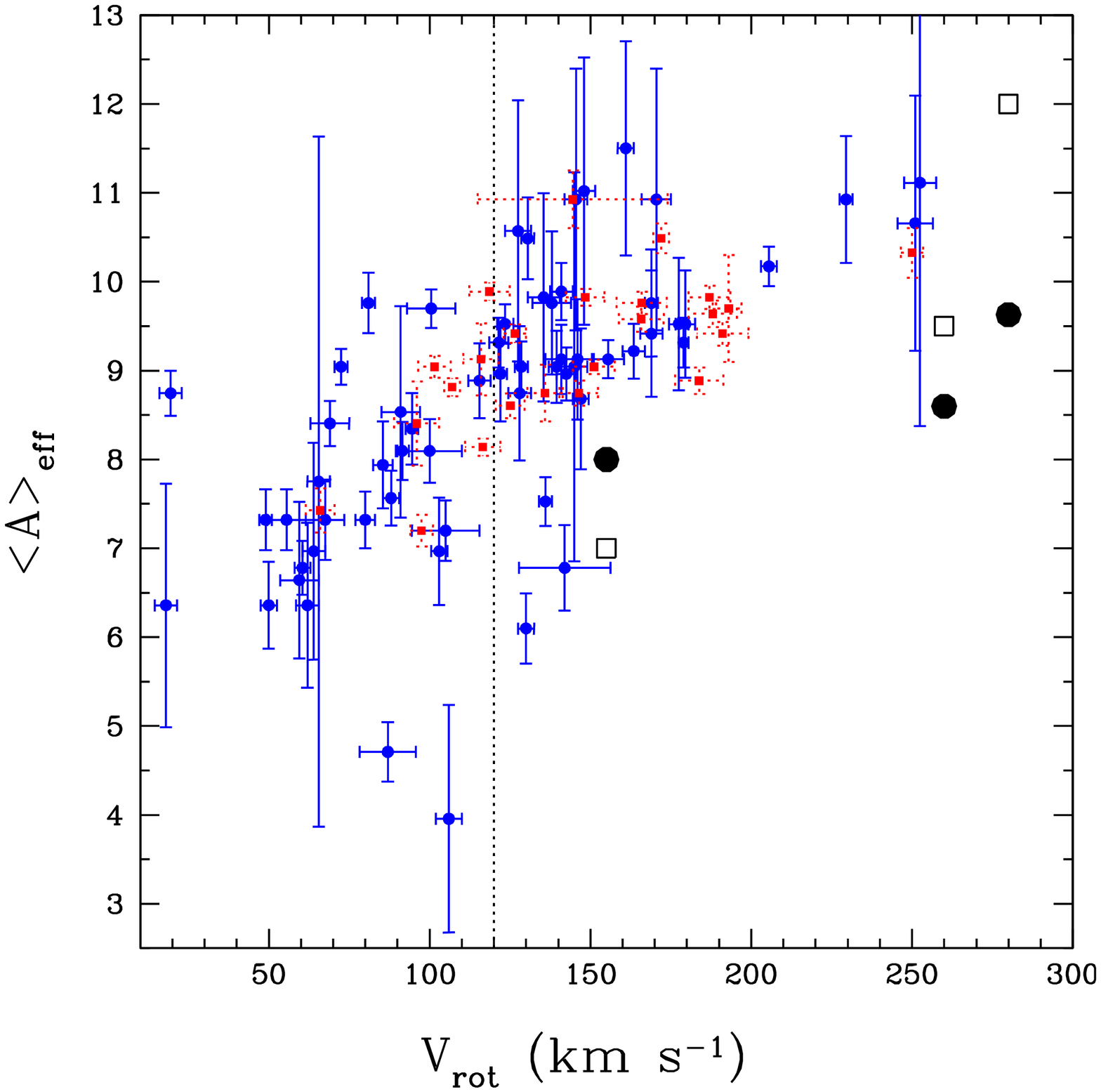}
\caption{Left panel: Star formation rate of all stars within 4 disk
scale heights from the disk plane in two different runs of DWF1. Black
solid line: strong SN feedback ($\epsilon$SN=0.6); Magenta
long-dashed: no feedback, no UV. The addition of feedback smooths out
the peak in SF otherwise present at high redshift and during the last
major merger event at z=2.3. Feedback delays the conversion of gas
into stars until gas accumulates and cools in the potential well of
the main progenitor.  The run with feedback has a present day SFR
almost ten times higher. Right panel: Average and median age of
stellar populations vs galaxy circular velocity. Solid and empty
circles show values for our simulated galaxies.  Data points with
error bars are from MacArthur et al 2004.}
\label{twobarrel}
\end{figure}

\section{Results}
 
 \noindent {\it \bf Disks and Star Formation Rates:} At the final
time, the three galaxies have formed a substantial disk component well
fit by an exponential profile outside the inner 1-2 kpc.  The stellar
disks follow exponential profiles out to 2.5--3 scale lengths.  All
three galaxies also have extended disks of cold gas with exponential
scale lengths larger than even the youngest component of the stellar
disks.  Beyond 3 scale lengths, the stellar disk declines rapidly even
for young stars. Such disk profiles are quite common in observed
spiral galaxies with 150 $< v_c <$ 250 km/sec (Pohlen \& Trujillo
2005).

Within the inner 1-2 kpc, the two smaller galaxies (DWF1 and MW1)
exhibit a steep central ``bulge'' component. Due to its small physical
size, it is possible that this bulge component is affected by
resolution.  We confirmed this possibility by comparing to a run of
MW1 that uses a force resolution of 0.3 kpc and eight times better
mass resolution (this resolution is unprecedented for SPH simulations
of individual galaxies in a cosmological context).  While this
ultra-high resolution run produces a galaxy with a very similar disk
scale length and rotation curve, the central concentration of baryons
within the central kpc (and thus the central steepening of the
profile) is reduced (see Governato et al 2005).
\begin{figure}[t]
\plottwo{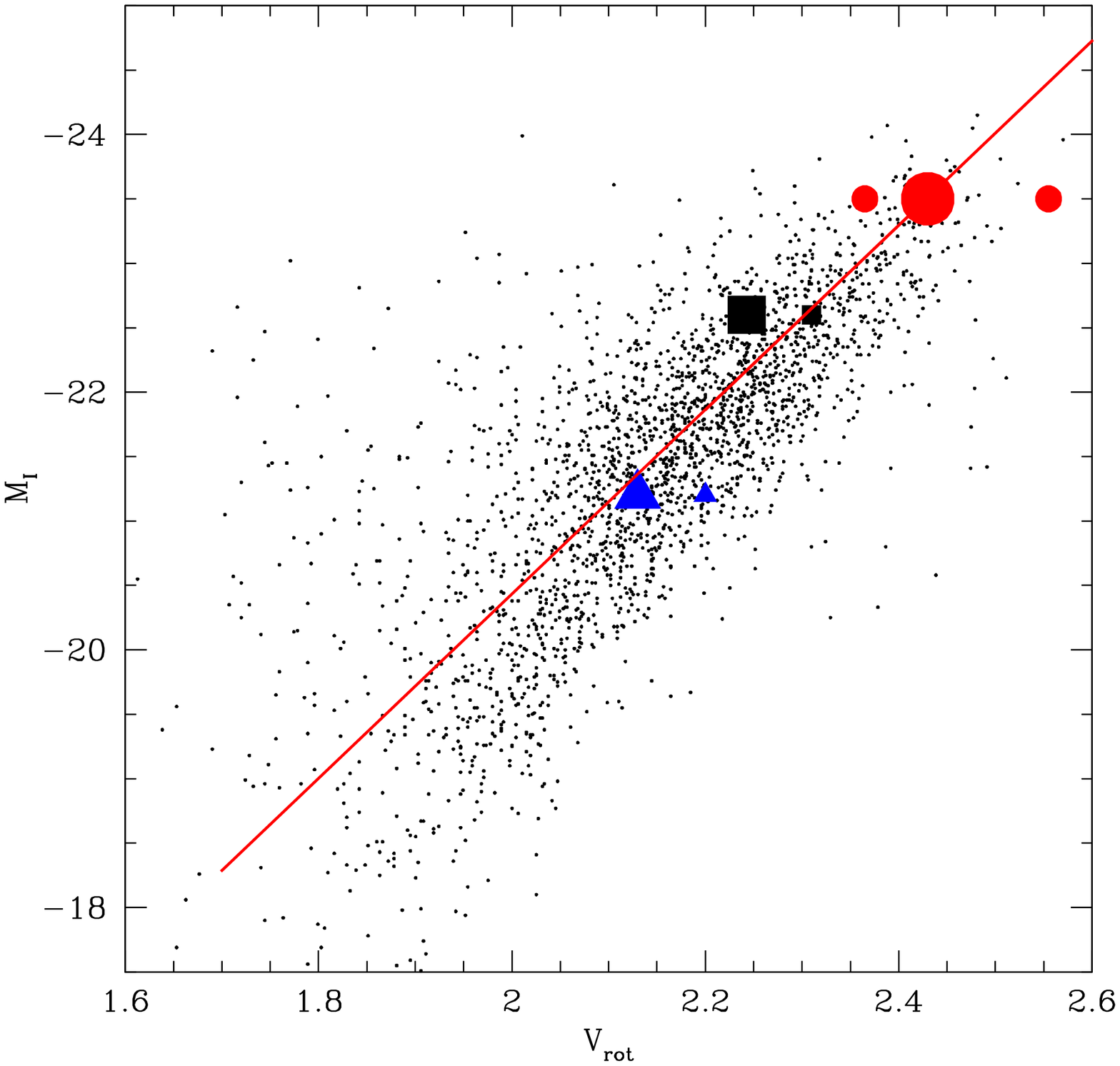}{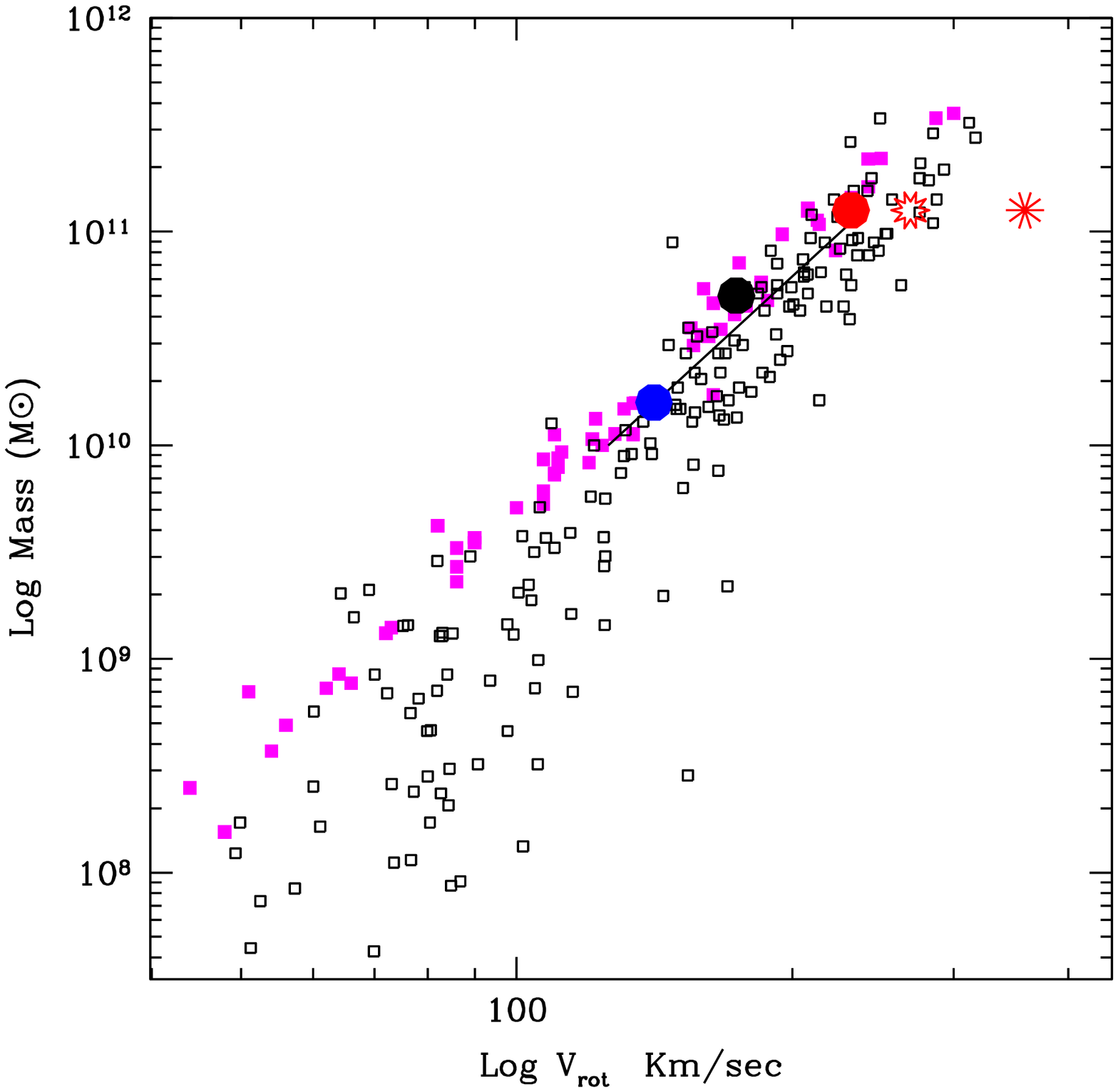}
 \caption{Left panel:  The Tully Fisher
relation.~Small dots are from a recent data compilation by Giovanelli
\& Haynes. The red line is a fit to data in Giovanelli et al
(1997). Solid triangle: DWF1; Solid Square: MW1; Solid/Starred Dot:
GAL1. Bigger dots shows V$_{rot}$ measured at 3.5R$_d$ (as in the
observational data).  Smaller dots show V$_{rot}$ measured at
2.5R$_d$. The rightmost red dot uses the stellar peak velocity. Right
Panel: The baryonic TF relation. Solid triangle: DWF1; Solid Square:
MW1; Solid Dot: GAL1; Empty circles: McGaugh et al. 2000, Magenta
squares: McGaugh et al 05. The solid line is a best fit from data in
Gurovich et al. 2004. Simulation data points include all stars and cold gas
within their bulges \& disks components.}
\label{twobarrel}
\end{figure}
\noindent We also investigated the effect of SN feedback
 strength on the final galaxy properties.  The stellar profile and
 size of the final disks in the more massive galaxies (MW1 and GAL1)
 is only weakly dependent on the strength of SN feedback because the
 adopted efficiency for supernovae heating is too low to generate
 large scale winds and mass loss in massive halos. The fact that the
 final stellar and cold gas masses compare very well with the Milky
 Way thus argues in favor of such relatively moderate effect of
 feedback at these mass scales. However, the dwarf galaxy disk stellar
 profile is drastically affected by varying amount of SN feedback.
 When $\epsilon$SN is increased to 0.6, the stellar density profile of
 DWF1 is best described by a single exponential profile with a shorter
 scale length than that measured with $\epsilon$SN = 0.4.  Unlike the
 disk sizes, the strength and longevity of dynamical bars in all three
 galaxies increases remarkably with decreasing strength of the
 feedback.  SN feedback also plays a major role in shaping the star
 formation histories (SFHs) of our simulated galaxies, particularly
 that of DWF1, the lowest mass galaxy in the set.  The right panel of
 Fig.~1 shows that the simulated z=0 SFRs are in good agreement with
 those measured from the SDSS over a large range of stellar masses
 (Brinchmann et al 2005).  The left panel of Fig.~2 compares the SFH of
 DWF1 in the cases without SN feedback and with the strongest feedback
 adopted in our study ($\epsilon$SN = 0.6). Early star formation is
 significantly reduced by feedback, as the relative strenght of SF
 bursts during major merger events.  We verified that lowering the
 efficiency of star formation during mergers preserves a large
 quantity of cold gas that rapidly settled in rotationally supported
 disks. Fig.~2 (right panel) also shows that our simulations follow the
 trend observed by MacArthur et al. (2004): galaxies with less massive
 stellar component also tend to have younger stellar populations.  The
 inclusion of feedback and its differential effect at different galaxy
 masses reproduces the observed ``anti-hierarchical'' trend within a
 cosmological context.

\noindent {\it \bf The Abundance of Milky Way Satellites:} A
 combination of radiative feedback from the cosmic UV background and
 SN winds drastically reduces the number of satellites within the
 virial radius of the MW1 galaxy to the point where it is in agreement
 with observations, although the satellites are too bright compared to
 those of the Milky Way. The satellites resolved in these runs
 ($V_{max}$ > 25 km/s) exhibit extended and sporadic star formation
 histories, similar to those observed in Milky Way dwarfs.

 \noindent {\it \bf Rotation curves and the Tully--Fisher relation:}
 The azimuthally averaged rotation curves of galaxies simulated
 including SN feedback are shown in the left panel of Fig.~1. This
 figure shows that gas rotational velocities are very similar to
 stellar ones for models with feedback. However, gas and stellar
 velocities are often different for models without feedback, because
 stars have higher velocity dispersions and cold gas is often
 completely misaligned with the main stellar component.  The rotation
 curves combined with the disk scale lengths of intermediate age disk
 stars allow us to compare our galaxy models against two fundamental
 relation for normal galaxies: (1) the Tully--Fisher (TF) relation
 that links the characteristic rotation velocity of a galaxy with its
 total magnitude or baryonic disk+bulge mass (Giovanelli \& Haynes
 1997) and (2) the ``baryonic TF'' relation (McGaugh 2002, McGaugh
 2005), which takes into account that smaller galaxies are more gas
 rich and that stars account for only a fraction of their disk total
 mass (Fig.~3).  Early simulations reported severe difficulties in
 matching the normalization of the above relations, as the central
 rotational velocities of the simulated galaxies were too high due to
 an excess of matter at the center of galaxies. This was possibly due
 to excessive baryon cooling and angular momentum loss and the
 subsequent adiabatic contraction of the DM component. High resolution
 simulations in which only a weak feedback was included, such as those
 of Governato et al (2004) and Abadi et al (2003), showed a small, but
 consistent shift from the TF relation with galaxies being too
 centrally concentrated compared to real ones.  Our new set of
 simulations successfully matches both relations. {\it Matching the
 Tully-Fisher relation is possibly the most important conclusion of
 the work presented here.} This match shows that our simulations
 produce a realistic distribution of stars, baryons and dark matter
 within the central few kpc of disk galaxies. Much progress remains to
 be made before galaxy formation is fully understood. However, once
 sufficient resolution and some of the complexities of star formation
 and SN feedback are introduced in numerical simulations, the
 $\Lambda$CDM scenario is able to create galaxies that share some of
 their structural properties with the real ones.

\begin{acknowledgments}

FG was a Brooks Fellow when this project started. FG was supported in
part by NSF grant AST-0098557 and the Spitzer  Telescope Theory
Grant. Simulations were run at ARSC. We thank our collaborators for
allowing us to show results before publication.

\end{acknowledgments}

%
% Bibliography made with BibTeX:
%% kapalike is preferred if you have used \kluwerbib, above.
%% Otherwise you may use any .bst style your editor approves.

%or 

\end{document}